\documentclass[onecolumn,12pt]{aastex701}
\usepackage{graphicx} 
\usepackage{wrapfig}
\usepackage{myusepacknojournals}

\newcommand{\fen}{$f^{e}_{900}$\/}

\newcommand{\flux}{ergs cm$^{-2}$ s$^{-1}$ \AA$^{-1}$}

\newcommand{\hwo}{{\em HWO}\/}

\begin{document}

\title{The short and long of iPhoton science for a boosted Hubble}

\author[0000-0003-0503-4667]{Stephan R. McCandliss}
\affiliation{Johns Hopkins University, Department of Physics \& Astronomy, Center for Astrophysical Sciences, 3400 North Charles Street, Baltimore, MD, USA, 21218}
\email[show]{stephan@pha.jhu.edu}

\author[0000-0002-1552-3600]{Jack Ford}
\affiliation{Johns Hopkins University, Department of Physics \& Astronomy, Center for Astrophysical Sciences, 3400 North Charles Street, Baltimore, MD, USA, 21218}
\email{jford51@jh.edu}

\author[0000-0002-6790-5125]{Anne E. Jaskot}
\affiliation{Department of Astronomy, Williams College, Williamstown, MA 01267, USA}
\email{08aej@williams.edu}

\author[0000-0001-8587-218X]{Matthew J. Hayes}
\affiliation{Stockholm University, Department of Astronomy and Oskar Klein Centre for Cosmoparticle Physics, AlbaNova University Centre, SE-10691, Stockholm, Sweden}
\email{matthew.hayes@astro.su.se}  

\author[0000-0001-8419-3062]{Alberto Saldana-Lopez}
\affiliation{Stockholm University, Department of Astronomy and Oskar Klein Centre for Cosmoparticle Physics, AlbaNova University Centre, SE-10691, Stockholm, Sweden}
\email{alberto.saldana-lopez@astro.su.se}

\author[0000-0002-6586-4446]{Alaina Henry}
\affiliation{Space Telescope Science Institute, Baltimore, MD 21218}
\email{ahenry@stsci.edu}

\author[0000-0001-6670-6370]{Timothy Heckman}
\affiliation{Johns Hopkins University, Department of Physics \& Astronomy, Center for Astrophysical Sciences, 3400 North Charles Street, Baltimore, MD, USA, 21218}
\email{theckma1@jhu.edu}

\author[0000-0002-0159-2613]{Sophia R. Flury}\email{sflury@roe.ac.uk}
\affiliation{Institute for Astronomy, University of Edinburgh, Royal Observatory, Edinburgh, EH9 3HJ, UK}

\author[0000-0002-9136-8876]{Claudia Scarlata}
\affiliation{Minnesota Institute for Astrophysics, University of Minnesota, Minneapolis, MN 55455, USA}
\email{mscarlat@umn.edu}

\author[0000-0003-4166-2855]{Cody Carr}
\affiliation{University of Michigan Department of Astronomy 1085 S. University Ann Arbor, MI 48109, USA}
\email{codycarr@umich.edu}

\date{ 22 May 2026 - a White Paper submitted to the STScI request for Building a Roadmap for Hubble science into the 2030s}


\begin{abstract}
    Boosting the {\em Hubble Space Telescope} (\hst) will provide unique opportunities to carry out precursor science for the {\em Habitable Worlds Observatory} (\hwo).  Chief among them are science cases for determining the properties of star forming galaxies that contribute to creating and sustaining the universe in a mostly ionized state. The farUV and nearUV spectroscopic capabilities of the Cosmic Origins Spectrograph (COS) and the Space Telescope Imaging Spectrograph (STIS) are unique and unlikely to be replicated in the near future. Here we describe the benefits of a deep panchromatic spectroscopic effort to answer questions concerning the shape of the rest frame ionizing radiation escaping from star forming galaxies at modest redshift, capture crucial missing spectral regions, and explore whether their star formation histories are truly similar to the LyC leakers responsible for initiating and later sustaining the mostly-ionized-state of the universe.  An observing program emphasizing multi-orbit observations, unencumbered by \hst\ orbit competition and freely accessible to the wider ionizing photon (iPhoton) community, will catalyze crowd-sourced answers to these questions and offer a lower operating cost price point. \\[-.3in]
    
\end{abstract}

\section{Low z observations of ionizing radiation inform the physics of escape}

The past decade has seen remarkable success in direct detection of rest-frame ionizing radiation (Lyman Continuum - LyC) escaping from compact star-forming galaxies ($r_{50}$  $<$ 500 pc) with high star formation rates (10 - 100 M$_{\odot}$ yr$^{- 1}$).  These detections were made using the low spectral resolution ($R\sim$ 2000) mode of COS G140L CENWAV800 \citep{Redwine2016} to efficiently gather a sample of 66 potential leakers, chosen to have an overlapping diversity of high [\ion{O}{3}]/[\ion{O}{2}] ratios (O32), UV slopes ($\beta$), and star formation rate surface densities ($\Sigma_{SFR}$)  \citep{Jaskot2018}.  They were combined with other COS G140L observations in the literature from \citep{Izotov2016a, Izotov2016b, Izotov2018a, Izotov2018b} to produce an 89 object database, spanning the redshift range of 0.22 $\leq z \leq$ 0.45, collectively referred to as the Low z Lyman Continuum Survey plus (LzLCS+). These data have become the gold standard for inferring the physics of \lyc\ escape at high redshift by \jwst, as detailed in the review by \citet{Jaskot2025}.

\begin{figure}
\centering
\includegraphics[width=.9\textwidth, viewport=.9in .49in 10in 5.6in]{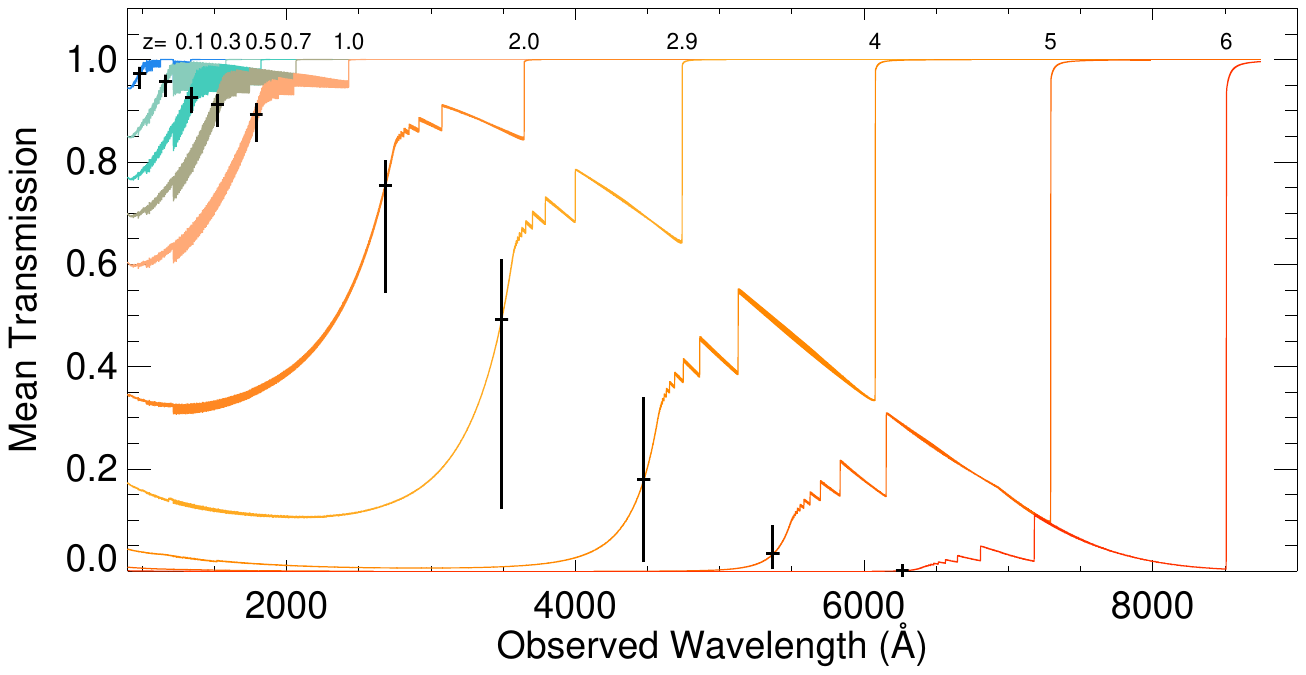}
\caption{\small  Mean IGM transmission  for  $z$ = (0.1, 0.3, 0.5, 0.7, 1.0, 2.0, 2.9, 4, 5, 6). Vertical bars mark the Lyman edge and indicate the level of expected variation found in a Monte Carlo study of IGM transmission \citep{Inoue2008}.}  \label{fig1}
\end{figure}

The emphasis on low-redshift observations as the laboratory for investigating the physical processes of LyC escape is driven by the mean opacity of the universe to ionizing radiation, as depicted in Figure~\ref{fig1}.  The opacity is caused by the overlap of neutral hydrogen clouds scattered throughout the universe, known as the Lyman $\alpha$ (\lya) forest, which becomes progressively thicker toward higher redshift.   The figure shows the result of integrating of Equation~\ref{eqtrans} over the distribution function of intergalactic neutral hydrogen absorbers per redshift interval $\left(\frac{\partial^{2}n}{\partial{N_{HI}}\partial{z}}\right)$, weighted by a total hydrogen and helium absorption profile \citep{Paresce1980, Madau1995, Inoue2014, McCandliss2017}.  

\begin{equation}
<\tau(\lambda)> = \sum_{i=0}^{L}\sum_{j=0}^{M} \left(\frac{\partial^{2}n}{\partial{N_{HI}}\partial{z}}\right)_{i,j} (1-e^{-\tau(\lambda_{o})} )\Delta z_i  \Delta N_j.
\label{eqtrans}
\end{equation}
We see that corrections for the mean opacity of the intervening forest progressively increase with increasing redshift becoming near impossible beyond $z \gtrsim$ 5.  Monte Carlo simulations of the forest by \citet{Inoue2008} show that the exact amount of attenuation on any given line-of-sight is highly variable, as indicated by the vertical crosses marking the observer location of the Lyman edge.  This leads to high uncertainty in the estimated \lyc\ flux at intermediate redshift and a reliance on “lucky sightlines” for direct detection \lyc\ leakage \citep{Steidel:2018}. The paucity of \lya\ absorbers at low redshift elevates the importance of the Lyman UV (LUV) band (defined as those wavelengths between \lya\ and the Lyman edge in the observer frame) for making direct detections of ionizing radiation escaping galaxies. In particular, the figure shows that at $z \sim$ 0.3  where much of the LzLCS+ data set was acquired, the intervening opacity is negligible.


\section{The short - the shape of ionizing radiation from star-forming galaxies}

\begin{figure}
\includegraphics[angle=90,width=0.46\textwidth]{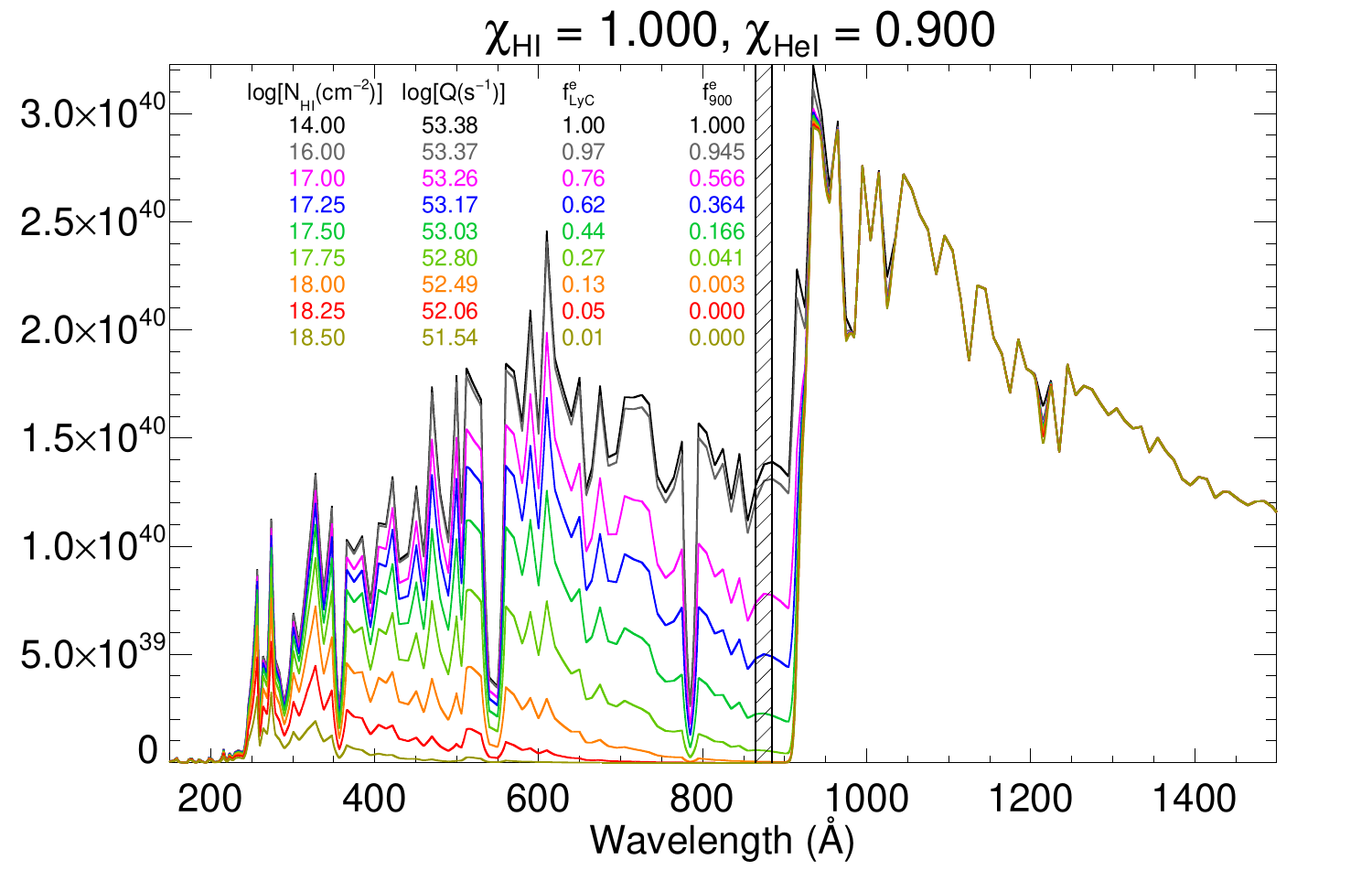}
\includegraphics[angle=90,width=0.5\textwidth]{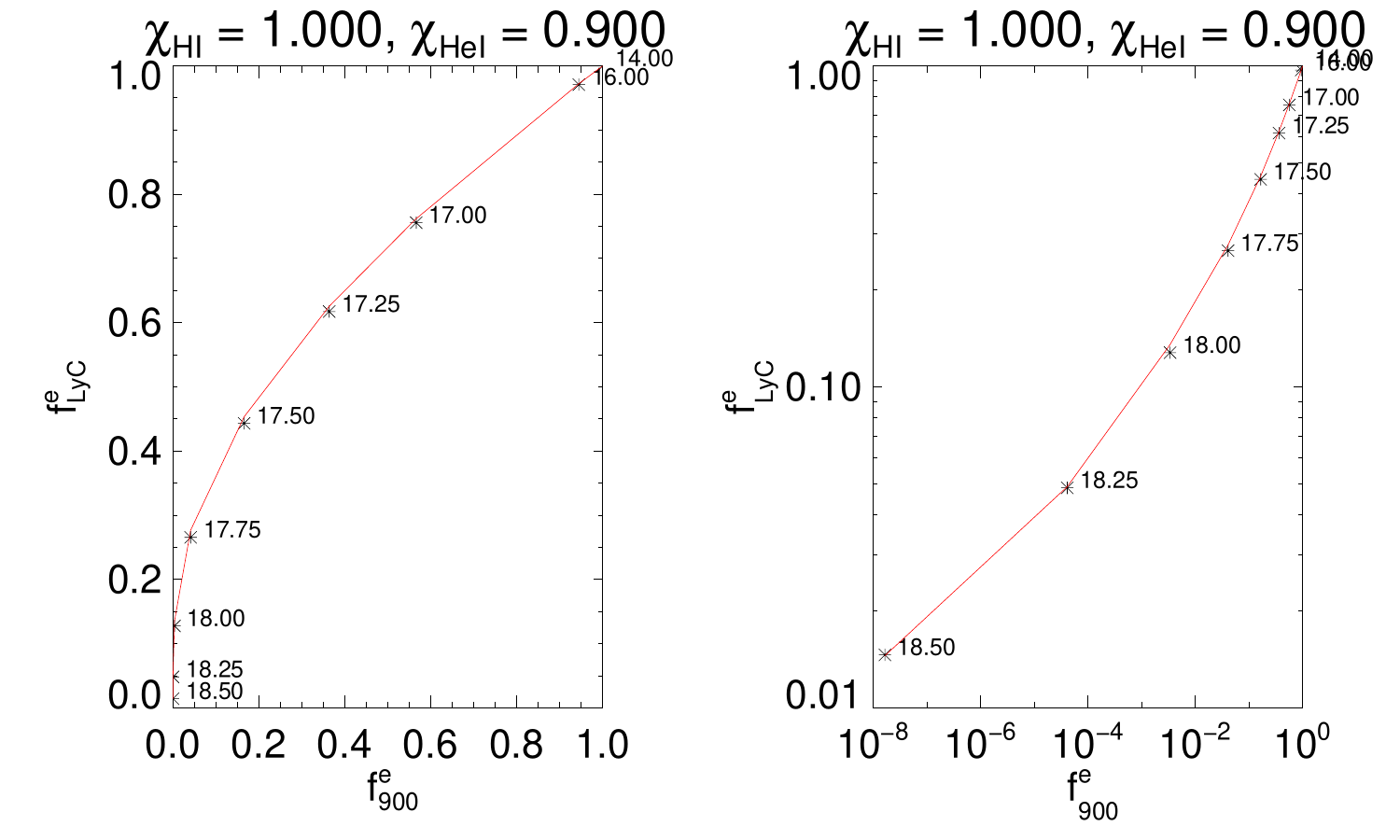}
   \caption{Left - Starburst99 \citep{Leitherer:2014} v0.4 evolution model (including stellar rotation - upper mass 100 $M_{\odot}$ - continuous star formation rate for 10 Myr) attenuated with the \ion{H}{1} ionization cross-section (Equation~\ref{hicross}). Right - integrated escape fraction \fec\ compared to the edge  escape fraction (\fen) in with linear and logarithmic axis. }
   \label{feint}
\end{figure}

Although large numbers of \lyc\ detections now exist, we have only recently begun to probe the actual shape of the LyC well below the Lyman edge.  These effects are important to account for because the escape of ionizing radiation measured just below the edge does not fully account for the total ionizing radiation budget integrated  over the whole \lyc\ emitting portion of a galaxy's Spectral Energy Distribution (SED). Elementary calculations show that it is possible for the ``edge'' escape fraction (\fen) to be very low (e.g. $< 1$\,\%), while the integrated \fec\ remains relatively high ($\sim 10$\,\%)  \citep{McCandliss2017}. This decidedly non-linear relationship is shown in Figure~\ref{feint}  

This naive calculation computes the ionizing radiation shape as the product of the intrinsic SED attenuated by the \ion{H}{1} absorption cross-section  (Equation~\ref{hicross}), 
\begin{equation}
\sigma_{\lambda} \approx \left\{ \begin{array}{ll}
6.3 \times 10^{-18} (\lambda/911.8)^{3} cm^{2}&\lambda \le 911.8\\
0&\lambda > 911.8 \\
\end{array}\right \} , 
\label{hicross}
\end{equation}
i.e. $F^{obs}_{\lambda} = F^{int}e^{-\tau_{\lambda}}$ with $\tau_{\lambda} = N_{HI}\sigma_\lambda$, where $N_{HI}$ is the neutral hydrogen column density. However,  calculations  \citep{Inoue2010, Inoue2011, Simmonds2024} and recent observations \citep[][Carr -- private communication]{Izotov2025} have shown that nebular emission in the form of free--bound transitions (free electrons recombining directly to $n =$ 1 level of hydrogen) will create a monotonically decreasing flux excess in a band $\approx$ 100~\AA\ wide shortward of the ionization edge at 911.8 \AA, dubbed the Lyman bump.   The presence of this Lyman bump makes it problematic to use the 850--911.8\,\AA\ region to determine the fraction of ionizing photons that escape from star-forming galaxies. The models suggest that the strength of this bump depends on the clumpiness and column density of the interstellar medium (ISM). By pairing LyC spectra with measurements of low-ionization species UV absorption lines, \hst\ can explore how \ion{H}{1} column densities, covering fractions, and density structures conspire to set the emergent LyC shape in different galaxies.

\begin{wrapfigure}[]{R}{.45\textwidth}
    \includegraphics[width=\linewidth]{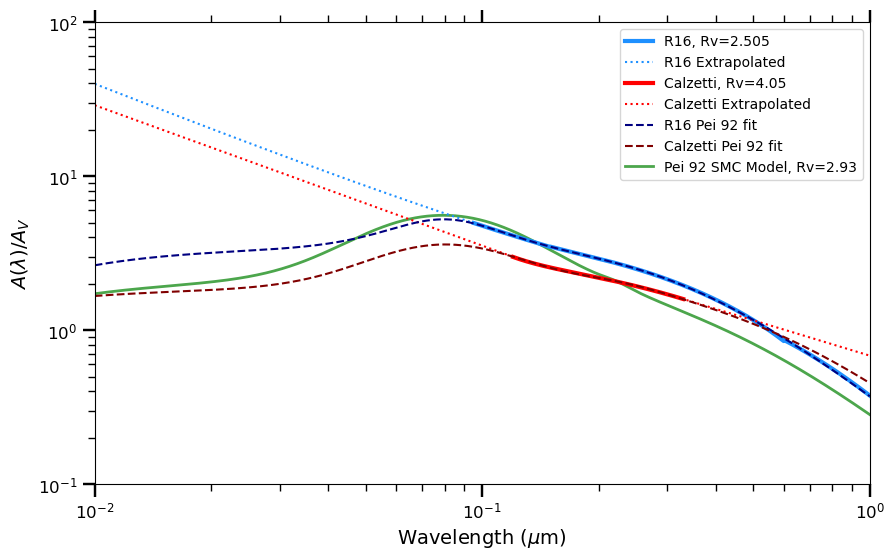}
   \caption{Dust models from \citet{Gordon2024} comparing simple extrapolations of empirical attenuation laws with theoretically guided extrapolations. }
    \label{dust}
\end{wrapfigure}

It is well known that dust generally ``reddens" a galaxy SED through absorption and scattering processes that increase with decreasing wavelength.   Variations in dust opacity beyond the Lyman edge are completely unexplored.  Theoretically, they produce attenuation softening in the ionizing continuum due to the real and imaginary parts of the dielectric functions adopted for graphite and amorphous silicates and are strongly dependent on  the size distribution and chemical composition of the astronomical dust \citep[c.f. the graphite/silicate models with the more recent Astrodust+PAH][]{Draine1984, Pei1992, Weingartner2001, Draine2003, Hensley2021, Hensley2023}.   Empirical models of dust attenuation as a function of wavelength \citep{Calzetti2001,Leitherer2002,Reddy2016,Reddy2018} are calibrated longward of the Lyman edge and offer little guidance beyond a simple extrapolation going shortward, although \citet{Gordon2024} has modified the empirical models to account for theoretical \lyc\ expectations over the edge as shown in Figure~\ref{dust}.


We propose using spectral synthesis models that account for star formation history and chemical enrichment in an unattenuated  "standard-galaxy" method -- similar to the standard-star method used to determine stellar extinction laws in the Milky-Way, Large and Small Magellanic Clouds -- to investigate and account for the effects of nebular emission, dust attenuation, and the continuous and discrete absorptions by \ion{H}{1} on the emergent shape of the star-forming galaxy SEDs both above and below Lyman edge. 

High resolution LUV spectra with sufficient continuum signal-to-noise (S/N $\sim$ 10) offer a wealth of high velocity stellar and low velocity ISM diagnostics in various low, medium and high ionization atomic species that are essential for assessing the chemical composition and outflow kinematics in the multiphase media of these objects. Low resolution spectra of objects with sufficient redshift  offer deep dives into the shape of the ionizing continuum.



\section{The long - filling in the missing wavelength coverage of LzLCS+}

\begin{wrapfigure}[]{R}{.55\textwidth}
    \centering
\includegraphics[width=\linewidth]{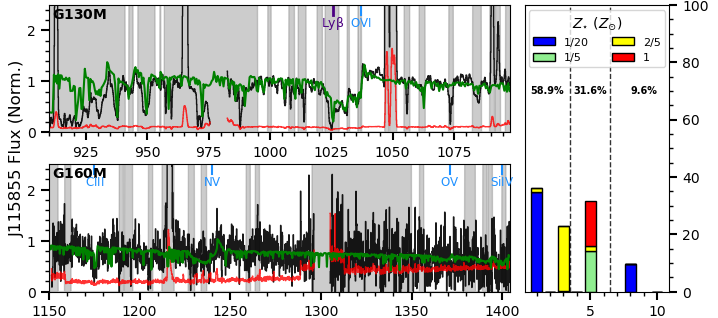}
\includegraphics[width=\linewidth]{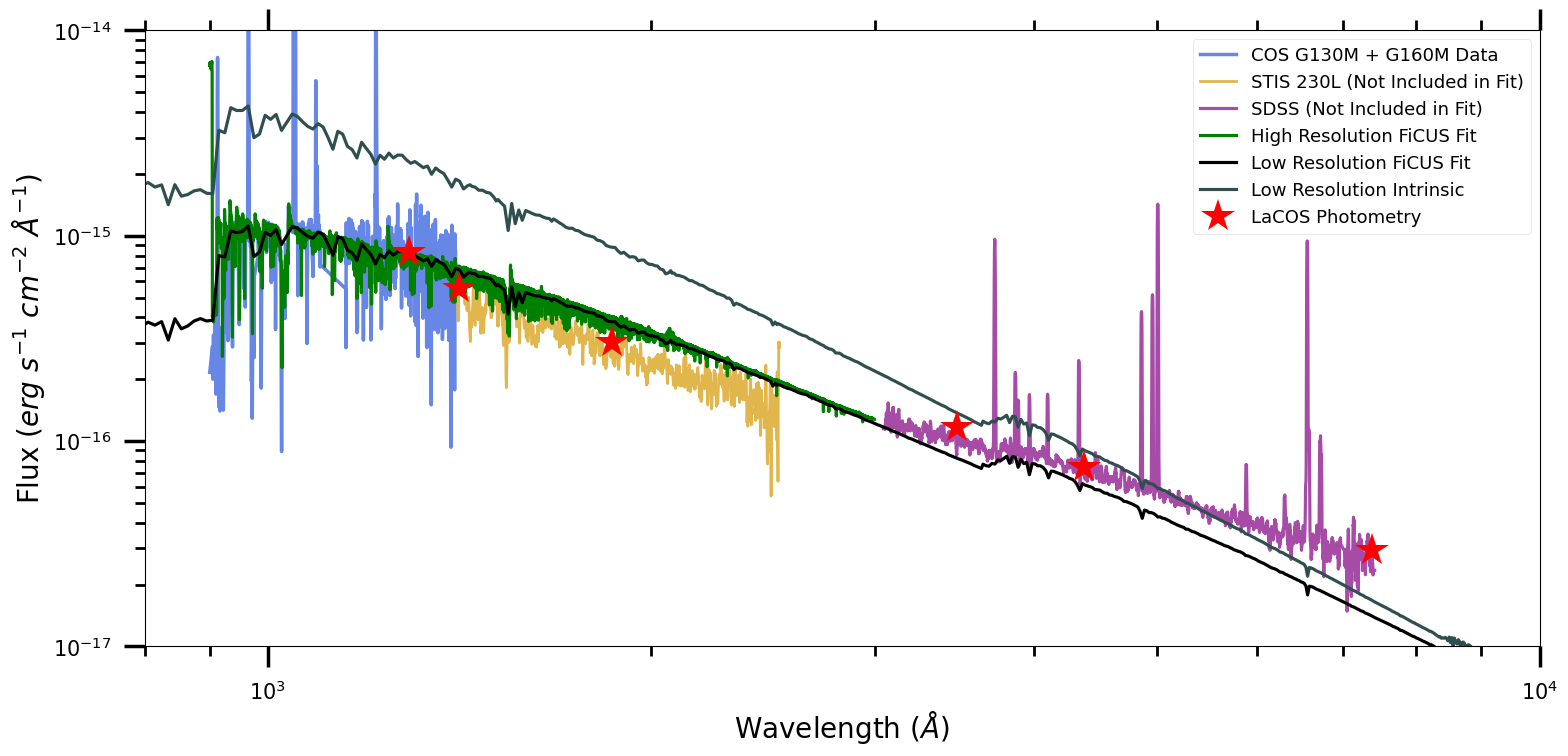}

    \caption{FICUS fit example for J115855+312559. Top left - fitted G130M and G160M range. Top right - light weighted age distributions. Bottom - comparison of fit to STIS and SDSS spectra.  Red stars are LaCOS photometery \citep{LeReste2025}.  }
    \label{ficus}
\end{wrapfigure}

A limitation of the LzLCS+ data set is the restricted wavelength range over which the spectra were acquired.  The G140L CENWAVE=800 observer frame bandpass did not allow observation of most restframe spectral features longward of the \ion{Si}{4} \dlam 1394,1403 doublet at the median redshift of $z \sim$ 0.3. This precludes including a more optimal range of ionization species in the fitting process, such as \ion{N}{4}] \dlam 1483, 1486, \ion{C}{4} \dlam 1548,1550, \ion{He}{2} \lam 1640, \ion{O}{3}, \dlam 1660,1667, [\ion{N}{4}] \lam 1719 and \ion{N}{3}] \lam 1750.  These lines can offer higher fidelity age and metallicity constraints on the star formation and chemical enrichment histories of the age-weighted spectral synthesis models \citep{Saldana-Lopez2022, Chisholm:2019} that we require to accurately assess the effects of dust, nebular emission on emergent SEDs. This wavelength region is particularly sensitive to determining the presence of evolved Wolf-Rayet (WR) populations that have been implicated in observations of enhanced nitrogen abundances at high redshift \citep[c.f.][]{Charbonnel2023,Senchyna2024,Schaerer2026}. The presence of a WR population is a key diagnostic for determining the age of the ionizing population as are the ancient massive stars that have evolve off the main sequence after $\sim$ 3 Myrs entering luminous blue variable stages and eventually supernova. Only the NUV modes in COS and STIS can access this wavelength interval.

Pushing observations of the LzLCS+ objects to visible and infrared wavelengths offers the opportunity to determine if older star populations are present that might be associated with extremely ancient star formation to address the question of whether these objects are true analogs of the star-forming galaxies at high $z$. It is reasonable to expect an analog of high $z$ ionizers (with low metallicity) to show signs of recent star formation up of $\la$ 1 Gyr -- the look back time at $z$ = 6. However, it is unreasonable for such objects to be considered analogs if they show signs of ancient star formation $>>$ 1 Gyr.

Figure~\ref{ficus} shows a FICUS (FItting Continuum of Ultraviolet Spectra; \citealt{Saldana-Lopez2023, Saldana-Lopez2025})  fitting example for the LzLCS+ target J115855+312559 with a set of single-age stellar population synthesis models up to 10 Myr.  The light weighted age favors objects $<$ 3 Myr.  It includes a meld of COS G130M, G160M, STIS G230L, and SDSS spectra.  We see it does a reasonable job of following the observed SED over COS far-UV portion of the combined spectra, but that it fails to match STIS and the SDSS, which may be due in part to incorrect aperture corrections.  However, this incarnation of FICUS only  only included the Starburst99 models up to 10 Myr, so the lack of agreement at the longer wavelength end is not surprising.  The code is now being updated to include much older stellar populations.  Nevertheless, a complete model of the star-formation history out to a few Gyrs will be required to establish the "analog or not" nature of the low-z leakers. 
 
\section{Going to Higher \MakeLowercase{z}}

\begin{figure}
    \centering
\includegraphics[width=.8\textwidth,clip,viewport=.8in 3.35in 8.2in 7.83in]{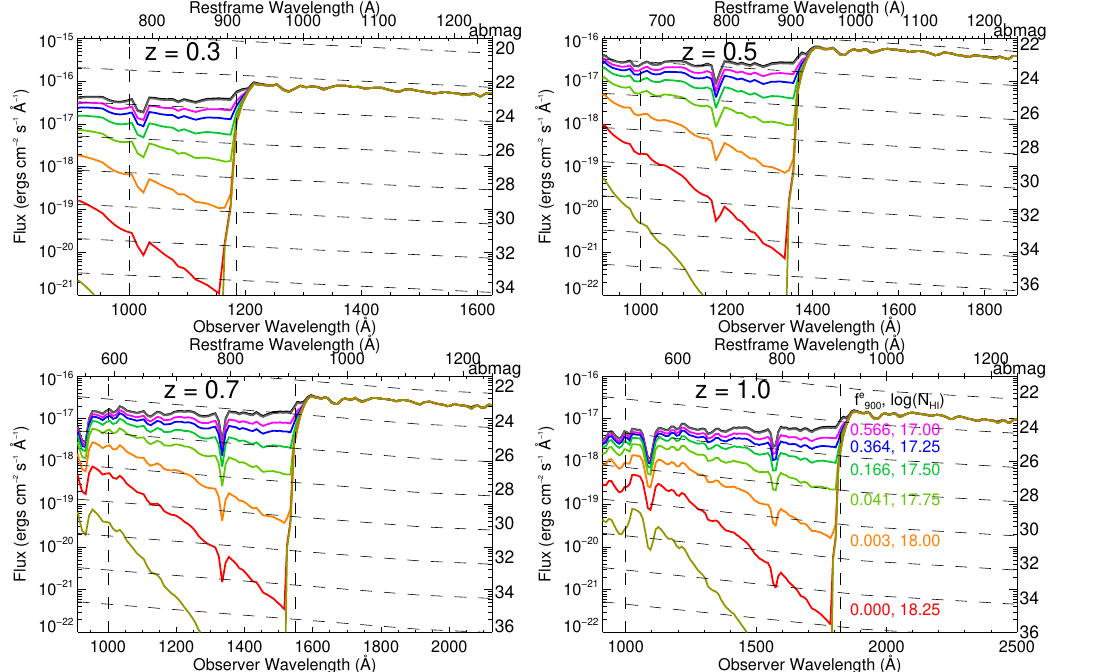}
    \caption{Redshifted attenuated Starburst99 models of $L^*_{1500(z+1)}$ galaxies with logarithmic scaling, showing LyC drop-ins towards shorter wavelengths. Contours of constant AB magnitude appear as dashed lines.  Escape fractions at 900 \AA\ ($f^e_{900}$ = 0.000, 0.000, 0.003, 0.041, 0.166, 0.364, 0.566, 0.945) correspond to column densities $\log{N_{HI}(cm^{-2})}$ = 18.50, 18.25, 18.00, 17.75, 17.50, 17.25, 17.00, 16.00); shown in olive, red, orange, light green, green, blue, violet, and grey respectively. The $z$ = 1 panel, middle-right, shows the  $f^e_{900}$ escape fractions and associated column densities.}
    \label{leakersnodust}
\end{figure}

All the LzLCS+ objects are very faint with observer frame fluxes close to the background limits of COS and STIS. Such objects require a long visits to obtain the requisite S/N. They were found by cross-catalog search between \galex, which supplied the FUV fluxes, and early SDSS spectroscopy data releases, which supplied the redshifts from the northern hemisphere. Since then the spectroscopic surveys offered by SDSS and the {\em Dark Energy Spectroscopic Instrument} (DESI) have expanded our ability to search for the extreme emission line galaxies most likely to be \lyc\ leakers into the southern hemisphere (see Le Reste WP). UVEX will extend  the FUV catalog \citep{Fucik2024}.  These data should greatly expand the pool of star-forming galaxies beyond $z >$ 0.3 to as high as $z$ = 1  -- half the age of the universe -- \citep[see][]{Izotov2025}, allowing for deep investigations into the ionizing radiation shape over a rest frame EUV bandpass of 500 - 900 \AA.  By targeting objects that appear to drop-out at 900 \AA\ we may discover that some drop-in towards 600 \AA\ as indicated by Figure~\ref{leakersnodust}, as long as dust truly levels off as Figure~\ref{dust} appears to indicate.

\section{\MakeLowercase{i}Photon Observing Strategies }

The data acquired in a dedicated Treasury style program (see James WP) along with the analysis techniques described here will become an invaluable \hwo\ resource for optimizing spectrograph design parameters and allow efficient detection of ISM rest frame outflow signatures in the LUV and leakage in EUV.  It will also stimulate the community into creating a vast tool chest of models that can maximize iPhoton science return from day one.

How orbitally expensive would such programs be? The baseline is set by the COS and STIS background limits, which are on the order of $\sim$ 10$^{-17}$ \flux. One goal would be to complete the NUV using STIS wavelength coverage of the 89 LzLCS+ objects with redshift between 0.22 $< z <$ 0.45, to realize S/N $\sim$ 10 in the observer frame continuum at 2310 \AA.  At the mean \galex\ NUV ab-magnitude flux of these objects  $m_{2310} \sim$ 21 ($F_{2310}$ = 8$\times$10$^{-17}$ \flux).  The STIS exposure time calculator yields $\sim$ 5 orbits per average object to achieve this S/N goal, yielding a $\sim$ 450 orbit program.

Another goal would be to expand the sample to medium redshifts (a MzLCS). Here we can use the fluxes probed by \citet{Izotov2025} to offer orbital allocation guidance. Assuming a mean continuum flux of $m_{1500} \sim$ 22 ($F_{1500}$ = 7.8$\times$10$^{-17}$ \flux - essentially  the flux as above) for redshift $z \sim 0.5$ and a continuum SN = 5 per G140L Resel at rest wavelength 1000 \AA\ (1500 \AA\ observer frame) we would require about 20 orbits.  This would yield a sample of about 20 to 40 objects for a 500 to 1000 orbit program.  The advancement of our understanding of the shape of the \lyc\ out to 700 \AA\ would be unprecedented.

\bibliographystyle{aasjournalv7}
\bibliography{kincov} 

\end{document}